\documentclass[%
 reprint,
superscriptaddress,
 amsmath,amssymb,
 aps,
 prl,
floatfix,
]{revtex4-2}

\usepackage{upgreek}
\usepackage[dvipsnames]{xcolor} 

\usepackage{graphicx}
\usepackage{dcolumn}
\usepackage{bm}

\newcommand{\acos}{\operatorname{acos}}
\newcommand{\F}{\operatorname{F}}

\begin{document}


\title{High-Energy Photon Generation from Self-Organized Plasma Cavities in Field-Enhanced Laser-Preplasma Interactions}

\author{P. Hadjisolomou}
\email{Prokopis.Hadjisolomou@eli-beams.eu}
\affiliation{ELI Beamlines Facility, Extreme Light Infrastructure ERIC, Za Radnicí 835, 25241 Dolní Břežany, Czech Republic}

\author{R. Shaisultanov}
\affiliation{ELI Beamlines Facility, Extreme Light Infrastructure ERIC, Za Radnicí 835, 25241 Dolní Břežany, Czech Republic}

\author{T. M. Jeong}
\affiliation{ELI Beamlines Facility, Extreme Light Infrastructure ERIC, Za Radnicí 835, 25241 Dolní Břežany, Czech Republic}

\author{C. P. Ridgers}
\affiliation{York Plasma Institute, Department of Physics, University of York, Heslington, York, North Yorkshire YO10 5DD, UK}

\author{S. V. Bulanov}
\affiliation{ELI Beamlines Facility, Extreme Light Infrastructure ERIC, Za Radnicí 835, 25241 Dolní Břežany, Czech Republic}

\date{\today}

\begin{abstract}
The interaction of an ultraintense Nd:glass laser pulse with a near-critical plasma self-organizes into a highly efficient $\gamma$-ray source. Three-dimensional particle-in-cell simulations demonstrate that relativistic self-focusing, aided by a self-generated electron cavity, enhances the laser intensity by more than an order of magnitude, driving the system into the radiation-reaction-dominated regime, i.e. one where the electrons lose a substantial amount of their energy as hard radiation. Peak photon emission occurs near $0.5$ times the relativistic critical density, with a $\gamma$-photon yield exceeding $20\%$ of the laser energy. Compared to Ti:Sa lasers of the same power, the longer duration of Nd:glass laser pulses leads to an order of magnitude increase in $\gamma$-photon number in the extreme conversion efficiency regime, making them particularly well-suited for photonuclear physics applications. These findings point to a robust and scalable mechanism for compact, ultra-bright $\gamma$-ray generation in the multi-petawatt regime.
\end{abstract}

\maketitle




Ultraintense laser-plasma interactions at multi-petawatt power levels are rapidly advancing \cite{1985_StricklandD} into regimes where radiation reaction and quantum electrodynamics (QED) effects play a central role in shaping plasma dynamics \cite{2012_DiPiazzaA, 2022_GonoskovA}. At intensities exceeding $10^{22}\,\mathrm{W\,cm^{-2}}$, electrons emit “high-energy” $\gamma$-photons with energy $>1\,\mathrm{MeV}$ (hereafter referred to simply as “photons” for brevity) via nonlinear Compton scattering, and the related radiation losses strongly influence their dynamics \cite{2012_NakamuraT, 2012_RidgersCP, 2018_NielF, 2020_BlackburnTG}. These interactions hold promise for producing compact, ultra-bright $\gamma$-ray sources \cite{2023_HadjisolomouP} relevant to photonuclear physics \cite{2000_LedinghamKW, 2004_NedorezovVG, 2008_MullerC}, laboratory astrophysics \cite{1992_ReesMJ, 2010_RuffiniR, 2015_BulanovSV, 2018_PhilippovAA, 2021_AharonianF}, and fundamental tests of strong-field QED \cite{2012_DiPiazzaA, 2023_MacLeodAJ, 2024_MirzaieM}. Here, brightness is the main constraint, and we demonstrate how it can be increased by an order of magnitude, thereby enabling these applications in the near term.

Previous studies have shown that in near-critical plasmas within the regime of relativistic transparency, a self-organized channel can be formed \cite{1999_PukhovA, 2015_ArefievAV}. When it comes to photon generation, efficient regimes were found by employing either tightly focused pulses \cite{2021_HadjisolomouP, 2023_SugimotoK} or idealized preformed channels \cite{2022_HadjisolomouP_b}. However, simplified experimental arrangements dictate the use of robust targets, typically accompanied by a preplasma profile \cite{2004_EsirkepovTZ, 2008_LezhninKV, 2020_Hadjisolomou}. The combined effects of relativistic self-focusing, cavity formation, and radiation-reaction feedback remain incompletely understood under realistic conditions involving kilojoule-class, moderately focused laser pulses of long duration, such as Nd:glass laser pulses \cite{2025_RusB}. This work demonstrates that high photon yield and strong laser-to-photon coupling can be achieved at intensities and configurations available with kilojoule-class systems, making experimental realization feasible without requiring tightly focused or ultra-short pulses.

As the laser pulse propagates through the near-critical plasma, it undergoes relativistic self-focusing due to intensity-dependent changes in the refractive index \cite{1962_AskaryanGA_b, 1964_ChiaoRY, 1965_KellyPL, 1987_SunGZ}. This process leads to the formation of a self-generated electron cavity; a low-density channel carved out by the ponderomotive force, which expels electrons from high-intensity regions, and radiation pressure, which transfers momentum to the plasma. A useful physical framework for understanding this initial stage of cavity formation is provided by a “snowplow” model \cite{1999_BulanovSV, 2006_SazegariV}, wherein the intense laser field effectively sweeps up plasma electrons and compresses them into a narrow density front. This compression generates a space-charge field that pulls ions outwards, setting up a co-moving double layer and initiating a collective plasma motion. The snowplow model captures the quasi-steady balance between the laser radiation pressure and plasma inertia, leading to a forward-moving electron-depleted channel. This process initiates cavity formation and continues to shape the plasma structure until instabilities and magnetic field driven plasma restructuring effects dominate.

In this Letter, we present three-dimensional (3D) particle-in-cell (PIC) simulations of a $10\,\mathrm{PW}$ Nd:glass laser pulse (of dimensionless field amplitude, \mbox{$a_0 = e \, E / (m_e \, c \, \omega) \approx 80$}, where $E$ is the laser electric field, $e$ is the elementary charge, $m_e$ is the electron mass, $c$ is the speed of light in vacuum and $\omega$ is the laser frequency) interacting with a (relativistically) near-critical plasma of a hyperbolic tangent density profile. We show that relativistic self-focusing and cavity formation enhances the laser field by more than an order of magnitude, entering the radiation-reaction dominated regime at approximately ten times lower laser intensity than expected \cite{2015_ZhangP}. Peak $\gamma$-ray emission occurs near $0.5\,n_{cr}$ (the relativistic critical electron density is $n_{cr}= \gamma \, \epsilon_0 \, m_e \, \omega^2 / e^2$, where $\epsilon_0$ is the vacuum permittivity and $\gamma \approx \sqrt{1+a_0^2/2}$ for a linearly polarized laser), with excess of $20\%$ of the laser energy converted into high-energy photons. These findings establish the first experimentally accessible mechanism for generating ultra-bright $\gamma$-ray sources using multi-petawatt laser systems.

We use the 3D QED-PIC code EPOCH \cite{2015_ArberTD} to simulate the interaction of a laser of $\mathcal{E}_{las} = 1500\,\mathrm{J}$ energy, $150\,\mathrm{fs}$ duration (at full-width-at-half-maximum - FWHM) with a near-critical plasma. A characteristic wavelength of $1.06\,\mathrm{\upmu m}$ for Nd:glass systems is used. The laser is linearly polarized and focused to a $10\,\mathrm{\upmu m}$ spot size (at FWHM), giving a vacuum peak intensity of $8.3 \times 10^{21}\,\mathrm{W\,cm^{-2}}$. The target consists of fully ionized plasma (mass to atomic number ratio of 2) with a hyperbolic tangent electron density profile of the form
\begin{equation}
n_e = \frac{n_{e0}}{2} \left[1+\tanh \left(\frac{x-x_0}{w_0}\right) \right] ,
\end{equation}
with optimal (for photon emission) $x_0=98.304 \, \mathrm{\upmu m}$, $w_0 = 30 \, \mathrm{\upmu m}$ and $n_{e0} = 3.28 \times 10^{28} \, \mathrm{m^{-3}} \approx 0.56 \, n_{cr}$; this density corresponds to Silica Aerogel \cite{2014_SachithanadamM, 2014_WongJCH, 2021_IswarS}. The laser focal spot coincides with the location where $n_e=0.5 \, n_{e0}$ (at $x=x_0$), with the laser pulse propagating along the x-axis and its electric field oscillating along the y-axis. The simulation box spans from $0\,\mathrm{\upmu m}$ to $196.608~\mathrm{\upmu m}$ in x-direction and from $-24.576\,\mathrm{\upmu m}$ to $24.576\,\mathrm{\upmu m}$ in the other two directions. The x-direction is resolved with $16\,\mathrm{nm}$ and the others with $128\,\mathrm{nm}$ cells, with each cell containing 2 macro-electrons and 2 macro-ions. The Gaussian laser pulse is initialized such that its peak reaches the left simulation boundary after a delay of three standard deviations, $\sigma$. The simulations use the QED \cite{2012_RidgersCP} and Higuera-Cary \cite{2017_HiqueraAV} EPOCH modules.

\begin{figure}
  \centering
  \includegraphics[width=0.9\linewidth]{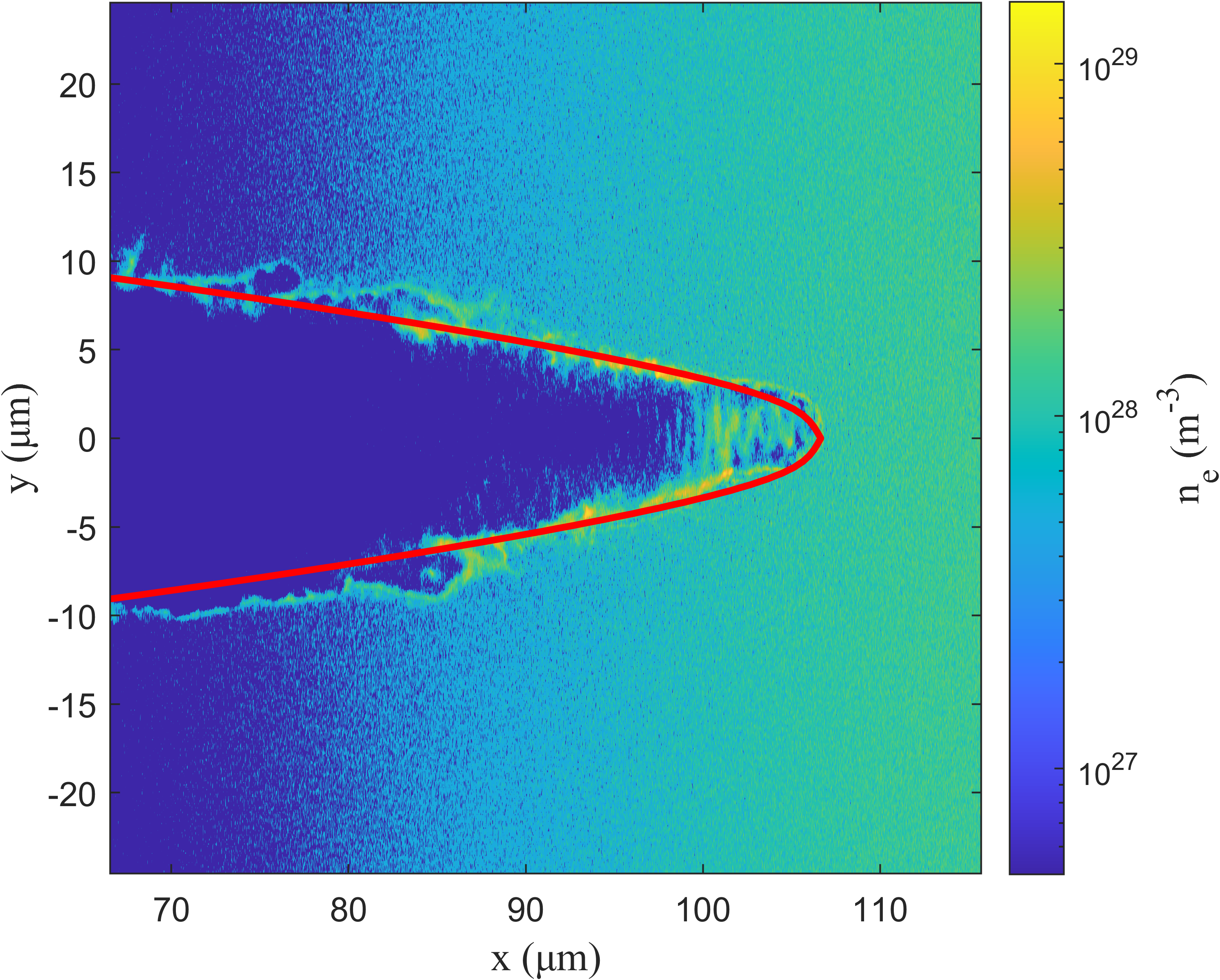}
  \caption{Electron number density in the $xy$-plane at $500\,\mathrm{fs}$ simulation time. The laser expels electrons transversely via the ponderomotive force, forming a self-organized, electron-depleted cavity that guides the laser over extended distances. The red line shows a fit to the cavity boundary using Eqs.~\mbox{\eqref{eq:two}--\eqref{eq:six}}, capturing the balance between laser radiation pressure and plasma response. This structure facilitates field intensification, thus enhancing photon emission.}
  \label{fig:1}
\end{figure}

Relativistic transparency enables partial penetration of the laser into the plasma. As the laser pulse sweeps electrons forward and expels them transversely, a strong longitudinal electrostatic field is established, enabling pre-acceleration and guiding conditions for electrons \cite{1999_PukhovA, 1999_GahnC}. Relativistic self-focusing leads to the formation of a narrow, electron-depleted channel. The early stages of this channel formation are consistent with the snowplow model, where the laser front accumulates and compresses electrons into a thin shell around an ion channel, establishing strong electrostatic fields that contribute to charge separation and ion motion. The expression for the overdense cavity boundary is given in Ref.~\cite{1999_BulanovSV}
\begin{align}
\frac{r}{\lambda^3} \sqrt{r^4 + s^4} 
&- \frac{1}{2} \F \left[ 
    \acos \left( \frac{r^2 - s^2}{r^2 + s^2} \right), 
    \frac{\sqrt{2}}{2} 
\right] \nonumber \\
&\quad + 3 \frac{s^2 \, c}{\lambda^3 v_g} (x - v_g \, t) 
= C ,
\label{eq:two}
\end{align}
where $r=\sqrt{y^2+z^2}$ is the cavity radius, $v_g$ is the group velocity, $t$ is the propagation time, $\F(u,\kappa)$ is the elliptic integral of the first kind, $s=r_0 \sqrt{p_0/(m_i \,c)}$, $m_i$ is the ion mass, $p_0$ is ion momentum, $r_0$ is the laser waist and $C$ is a constant. For linear polarization we have $s = r_0 \sqrt{ a_i / \sqrt{2}}$, where $a_i = a_0 \, m_e/m_i$. Since at the cavity front $x = v_g \, t$ and $r = 0$, then $C = \frac{1}{2} \F \left( \pi, \frac{\sqrt{2}}{2} \right)$.

However, self-focusing plays an important role in cavity evolution, which we acknowledge by modifying
\begin{equation}
s(x) = s_0 \, r(x) \, f(x)^{1/4} \, a(x)^{1/2} ,
\end{equation}
where $s_0 = 3$ serves as a fitting coefficient. Initially, we assume that self-focusing produces a truncated conical cavity, of the smaller cone diameter equal $\lambda$. Thus,
\begin{equation}
r(x) = r_0 - (r_0 - \lambda) \frac{x}{v_g \, t} ,
\end{equation}
Secondly, by assuming the pulse is long enough and that the power flow through two planes orthogonal to the propagation direction is conserved, we add a second correction term for field intensification within the cone, as
\begin{equation}
a(x) = \frac{a_i }{\sqrt{2}} \frac{r_0}{r(x)} .
\end{equation}
Thirdly, we acknowledge reduction of the pulse energy during cavity formation, through fitting to the numerical data for energy,
\begin{equation}
f(x) = \mathcal{E}_{min}/\mathcal{E}_{las}+\frac{ 1 -\mathcal{E}_{min}/\mathcal{E}_{las} }{1+ \exp[(x-x_c)/d] } ,
\label{eq:six}
\end{equation}
with fitting coefficients $\mathcal{E}_{min} \approx 141 \, \mathrm{J}$, $x_c \approx 116 \, \mathrm{\upmu m}$ and $d \approx 17.7 \, \mathrm{\upmu m}$. This coherent pushing action results in the initial shaping of the cavity before transverse and longitudinal instabilities fully develop. The formation of this cavity is visible in \mbox{Fig.~\ref{fig:1}}, where the fitted cavity walls are represented by the red line.

\begin{figure}
  \centering
  \includegraphics[width=0.9\linewidth]{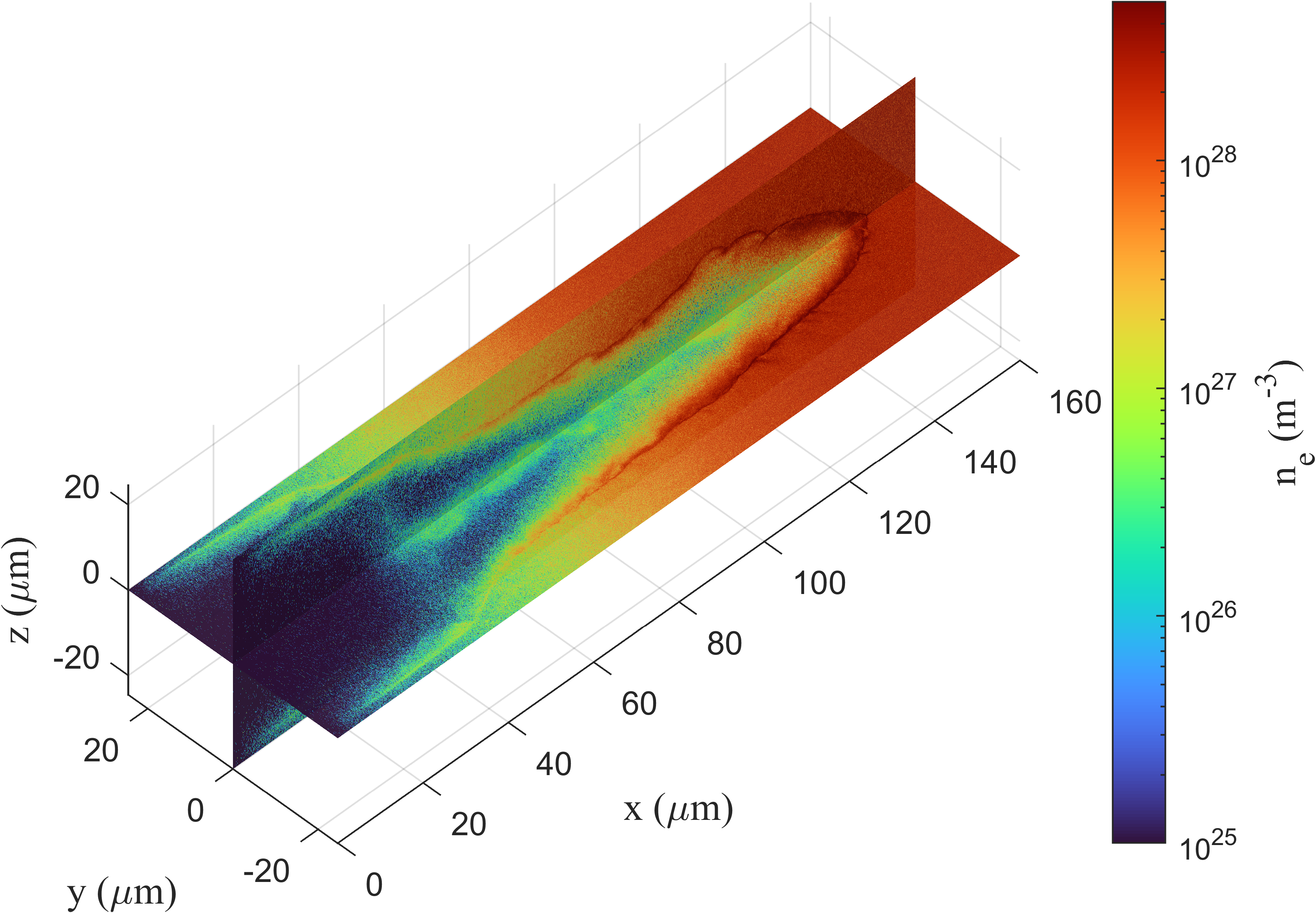}
  \caption{Electron number density at the end of the simulation. Prominent feature is the long-lived central filament.}
  \label{fig:2b}
\end{figure}

The cavity confines the laser energy, enhancing $a_0$ from an initial $a_0 \approx 80$ (in vacuum) up to $a_0 \approx 400$ (within the plasma), corresponding to an intensity enhancement of 25 times. However, as electrons reach ultra-relativistic energies, radiation reaction alters their trajectories, reduces dephasing length, and suppresses further energy gain \cite{2008_YorkAG}; ultimately limiting electron maximum energy, $\mathcal{E}_{\mathrm{max}}$, to a few GeV despite the extreme field strength. This interplay between snowplow-driven cavity formation and electron acceleration under strong radiation damping governs the electron dynamics and, by extension, the photon emission spectrum. In the high-electron-energy limit of near-vacuum laser electron acceleration, the maximum electron energy gain is governed by the scaling $\mathcal{E}_{\mathrm{max}} \approx 31 \sqrt{P\,(\mathrm{TW})} \, \mathrm{MeV}$, where $P$ is the laser power \cite{1995_EsareyE}. For a laser power of $P = 10\,\mathrm{PW}$, this yields an upper-bound energy gain of approximately $\mathcal{E}_{\mathrm{max}} \approx 3.1\,\mathrm{GeV}$. This estimate represents the ideal scenario in which electrons are injected with sufficiently high energy to avoid strong phase slippage, allowing sustained axial acceleration. Our QED-PIC simulations under these conditions yield electrons of $\sim 1.5\,\mathrm{GeV}$, indicating a reasonably efficient coupling in the ultraintense regime.

Charge separation fields and return currents play non-negligible roles in shaping the beam–plasma dynamics \cite{2004_PassoniM}. As electrons are driven forward and expelled radially by the laser field, strong electrostatic sheath fields develop along the channel boundaries. These fields increase sharply and can reflect lower-energy electrons, triggering photon emission in the mid-energy range (tens of MeV). Enhancing that photon population is particularly significant for inducing photonuclear reactions, as their cross-section peaks within that energy range.

The onset of electron motion in near-critical plasmas coincides with the development of strong azimuthal magnetic fields generated by transverse current imbalances \cite{1999_EsirkepovTZ, 2019_ParkJ}. These fields, reaching tens of kilotesla, exert a pinching force on the electron distribution, enhancing confinement within the cavity and eventually form a low-density filament along the laser propagation axis, behind the main pulse \cite{1999_PukhovA, 2021_JiangK, 2024_ValentaP}, as seen in \mbox{Fig.~\ref{fig:2b}}. In some simulations, we observe that in regions where hosing breaks the cavity symmetry, the magnetic topology becomes increasingly asymmetric, leading to randomized electron filamentation. Electron filamentation is linked to ion filamentation in the center of the cavity, observed in proton probing experiments \cite{2011_WillingaleL}.

\begin{figure}
  \centering
  \includegraphics[width=0.9\linewidth]{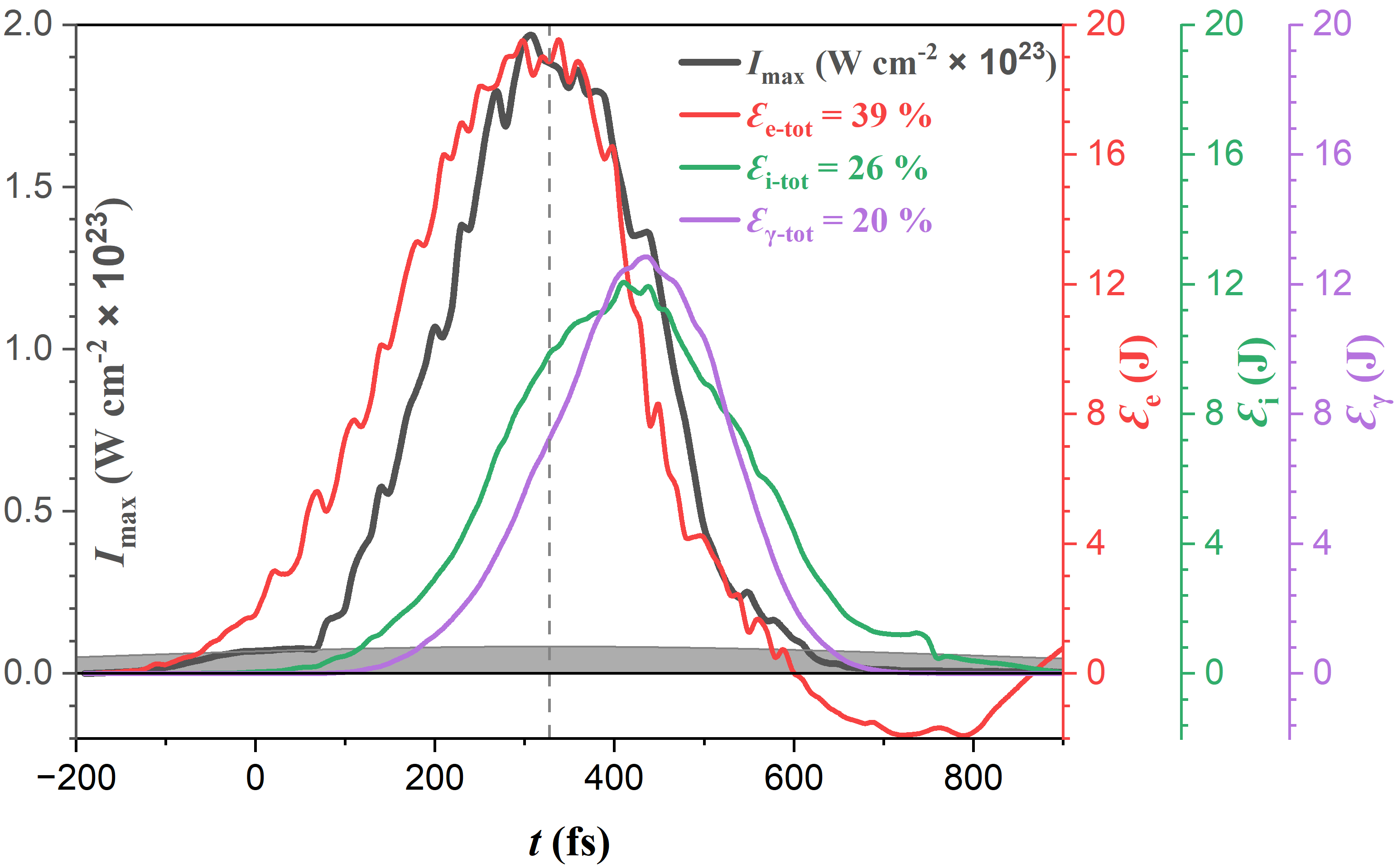}
  \caption{Left axis: The peak field intensity in plasma (black solid line) and in vacuum (black shaded area) vs time (shifted by $3 \, \sigma$). Right axis: The instantaneous laser to particle species conversion efficiency vs time. The red, green, and violet lines correspond to electrons, ions and photons, respectively; the integrated value is given on the label.}
  \label{fig:2}
\end{figure}

Our simulations reveal formation of distinct filamentary structures of smaller scale (comparable to the skin depth) ahead of the cavity boundary due to the fast electron population \cite{2004_RenC, 2006_TzoufrasM, 2012_BulanovSV}, characteristic of the Weibel-like current filamentation instability. These filaments emerge transversely to the laser propagation axis. The filamentation becomes more prominent at the plane of the magnetic field oscillation. This behavior is consistent with recent experimental observations using long-wavelength lasers \cite{2025_DoverNP}, and highlights the inherently collisionless nature of energy transport and instability growth in such regimes.

At around $0.5 \, n_c$, a near-symmetric well-defined cavity forms on the xy-plane, as shown on Fig.~\ref{fig:1}. On the xz-plane, symmetry occurs along the laser propagation axis, but accompanied with formation of secondary distinct cavities, oriented roughly at $20^\circ$. In agreement with our simulation data, experimental observations show that often the cavity initially distorts due to partial filamentation, but quickly self-corrects \cite{2011_WillingaleL}, as evident by the extended cavity formed at the end of the simulation, seen in \mbox{Fig.~\ref{fig:2b}}. The stable cavity formation results to the strong field amplification observed in our simulations and, consequently, peak $\gamma$-ray emission.

\begin{figure}
  \centering
  \includegraphics[width=0.9\linewidth]{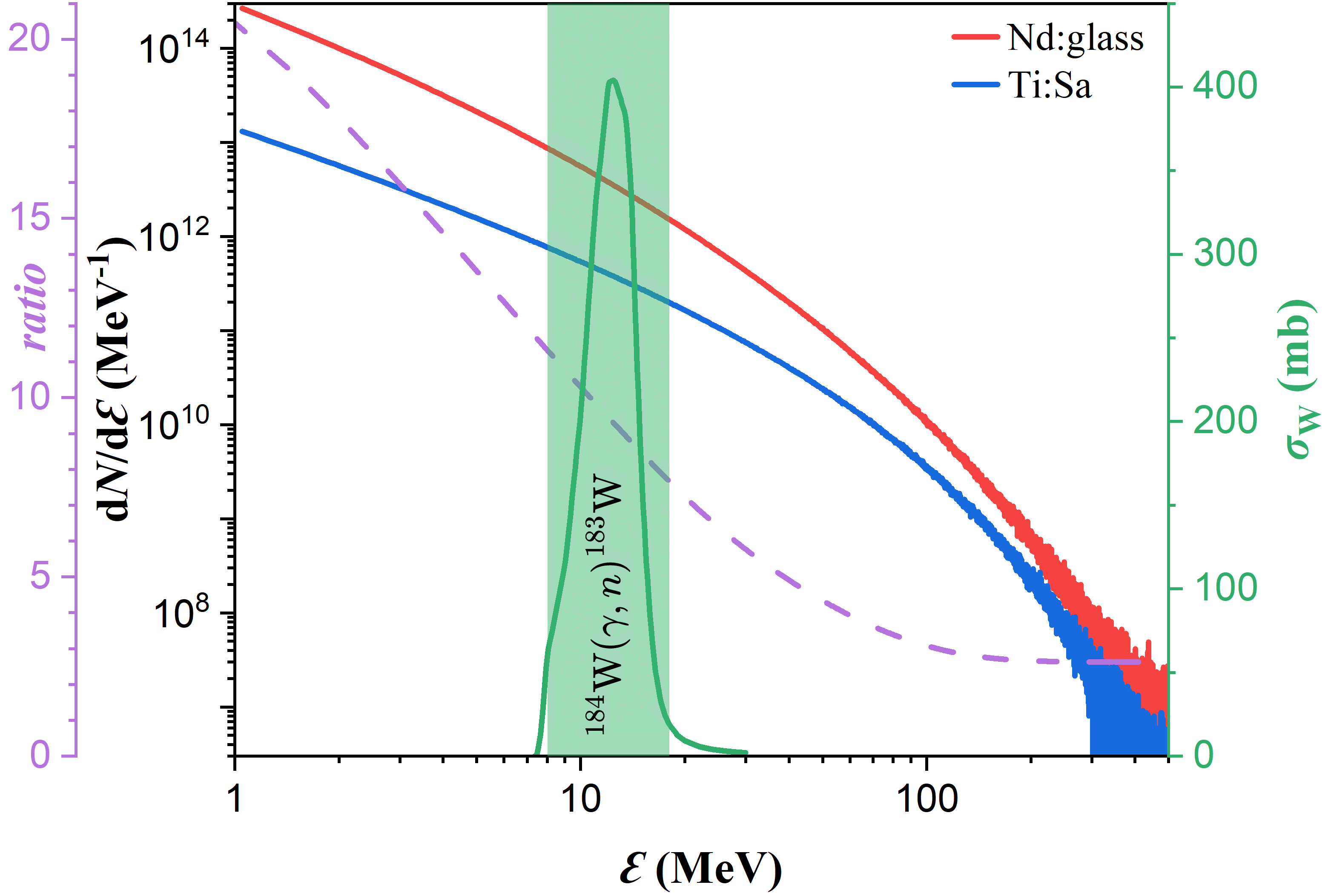}
  \caption{Energy spectra of emitted photons from a 3D PIC simulation of an ultraintense laser interacting with a near-critical plasma. The red line corresponds to the present simulation using an Nd:glass laser system. The blue line compares results from a previous study using a Ti:Sa laser system. Both lasers compared have a power of $10 \, \mathrm{PW}$. A broad distribution extending into the multi-MeV range is observed for both cases, but with the Nd:glass case yielding an order of magnitude more photons, as noted by the dashed purple line on the figure. The green line, highlighted by the green shaded region, indicates $\sigma_W$ for the $^{184}\mathrm{W}(\gamma,n)^{183}\mathrm{W}$ reaction.}
  \label{fig:3}
\end{figure}

Angularly resolved photon spectra reveal two predominant emission lobes, centred at $\pm 20^\circ$ with respect to the laser propagation axis and each lobe corresponding to an approximate cone of $10^\circ$ half-angle divergence. The evolution of laser intensity and photon conversion efficiency over time are shown by the black and red lines in \mbox{Fig.~\ref{fig:2}}, respectively. The photon energy spectrum extends up to $500 \, \mathrm{MeV}$, as shown in \mbox{Fig.~\ref{fig:3}}. The photon spectrum above $1 \, \mathrm{MeV}$ is well-fitted by
\begin{equation}
A_0 \, \exp \left( - \frac{\mathcal{E}}{\mathcal{E}_1} \right) \, \left[ 1 + \left( \frac{\mathcal{E}}{\mathcal{E}_2} \right)^{b} \right]^{-k} ,
\end{equation}
with $A_0=1.6 \times 10^{17} \, \mathrm{MeV^{-1}}$, $\mathcal{E}_1=110 \, \mathrm{MeV}$, $\mathcal{E}_2=100 \, \mathrm{MeV}$, $k=22.5$ and $b=0.243$. The photon spectra of a previous study \cite{2022_HadjisolomouP_b} using a Ti:Sa laser of $10 \, \mathrm{PW}$ power is compared by the blue line on \mbox{Fig.~\ref{fig:3}}. Both laser systems deliver $10,\, \mathrm{PW}$ of peak power, but differ in duration. While tighter focusing in the Ti:Sa case leads to higher initial intensities, the longer pulse duration and higher energy of the Nd:glass laser support extended interaction and stronger self-focusing over the pulse length. This leads to an order of magnitude more photons being produced in the Nd:glass case, with comparable spectral cut-off energy, as shown by the ratio of the two spectra (dashed purple line). These results underline the suitability of kilojoule-class, longer-duration pulses for driving cavity-enhanced photon generation at high efficiency.

A laser in an under-critical plasma can excite a plasma wave that travels along the laser pulse, resulting in field self-modulation \cite{1992_SprangleP, 1994_SprangleP}. Moreover, when a laser is self-guided in a channel it can undergo a hosing instability. This instability results from a displacement of the centroid, causing the ponderomotive force to kink the plasma cavity. Since the laser is guided into lower density regions, the misaligned plasma cavity enhances the laser centroid displacement, resulting in hosing instability. Thus, the propagation path of the laser is distorted and the cavity is modulated \cite{2016_CeurvorstL}, thereby affecting electron acceleration and, consequently, the spatial characteristics of photon emission.

We examined the parametric behavior of the density profile on hosing instability  by varying the amplitude, $n_{e0}$, and scale length, $w$, of the electron density profile. We find that the $\gamma$-ray yield reaches its highest values for $0.3<n_{e0}<0.6$ and $30<w_0<50$. Within this parameter range, although strong photon yield occurs, no strict trend emerges, likely due to the stochastic nature of the hosing instability. We observe that slower gradients (larger $w$) along with lower $n_{e0}$ values tend to enhance the growth of hosing and filamentation instabilities, as the laser pulse propagates further before dissipating its energy. On the other hand, steep gradients along with overly high $n_{e0}$ suppress effective electron acceleration, thereby limiting photon emission. These results indicate that precise control over the preplasma profile provides a valuable lever for optimizing both photon yield and beam quality, with direct implications for experimental design on upcoming multi-petawatt laser platforms.

Observation of side-scattered photons can provide a diagnostic of the plasma structure.  In the case with no hosing instability $\pm 20^\circ$ cones are observed. The growth of transverse instabilities such as hosing and filamentation leaves clear imprints on the angular and spectral photon emission patterns. These emissions are modulated by the evolving cavity that directs the laser field, suggesting a pathway for using radiation signatures as probes of nonlinear laser-plasma dynamics. The spatiotemporal evolution of photon hotspots correlates strongly with the onset of channel deformation, indicating that diagnostic imaging of $\gamma$-ray profiles may serve as a sensitive tool for studying instability growth in near-critical-density plasmas.

To quantify the brightness and applicability of the emitted radiation, we compute the peak brilliance of the source at an emission angle of $20^\circ$. The source duration is $230 \, \mathrm{fs}$ at FWHM, and the source size is approximately $1.4 \, \mathrm{\upmu m}$ (assumed as the region of the intensity FWHM at peak emission time). Thus, at $10 \, \mathrm{MeV}$ photon energy the brightness is approximately $8 \times 10^{22} \, \mathrm{photons/(s\,mm^2\,mrad^2\,0.1\%\,BW)}$, higher than proposed $\gamma$-ray emission schemes based on Ti:Sa lasers \cite{2022_HadjisolomouP_b}. Our simulations indicate that a single laser shot can yield more than $4 \times 10^{13}$ photons above $10\,\mathrm{MeV}$, positioning this mechanism advantageously for inducing photonuclear reactions, whose cross sections peak around that energy. Importantly, the radiation is emitted from a micron-scale volume, allowing tight spatial localization and enabling compact experimental geometries for photonuclear studies \cite{2023_LanHY, 2023_WuD, 2025_RasulovaFA}.

To assess the feasibility for nuclear applications, we estimate the yield of photonuclear reactions induced by the emitted $\gamma$-ray pulses. We adopt known cross sections (averaged to $\langle \sigma_W \rangle = 200 \, \mathrm{mbarn}$ for photons of $8-18 \, \mathrm{MeV}$, highlighted by the green area on  \mbox{Fig.~\ref{fig:3}}) for the $^{184}\mathrm{W}(\gamma,n)^{183}\mathrm{W}$ reaction \cite{2019_KoningAJ} and assume a tungsten slab of thickness $l=1 \, \mathrm{cm}$. We calculate the tungsten number density, $n_W=N_A \, \rho_W /A \approx 6.3 \times 10^{22} \, \mathrm{cm^{-3}}$, where $N_A$ is Avogadro number, $\rho_W \approx 19.25 \, \mathrm{g \, cm^{-3}}$ is tungsten mass density and $A \approx 183.8 \, \mathrm{g \, mol^{-1}}$ is tungsten atomic mass. By implementing $N_n = N_{\gamma} \, n_W \, \langle \sigma_W \rangle \, l$, we find that a single shot containing $N_{\gamma} = 3.8 \times 10^{13}$ photons of $8-18 \, \mathrm{MeV}$, can generate more than $4.8 \times 10^{11}$ photoneutrons in a compact secondary target. This opens opportunities for laser-driven nuclear diagnostics, isotope production, or ultrafast neutron radiography \cite{2013_RothM}. Additionally, the ultrashort duration and high energy of the source are attractive for time-resolved nuclear spectroscopy and probing QED pair creation thresholds in future setups with added background fields \cite{2012_DiPiazzaA, 2017_GonoskovA}.

In summary, we have demonstrated that kilojoule-class, moderately focused Nd:glass laser pulses interacting with near-critical plasmas can generate self-organized electron cavities that significantly enhance $\gamma$-ray production. The 3D simulations reveal that relativistic self-focusing and cavity formation increases the local laser amplitude by more than an order of magnitude, pushing the system into the strong radiation reaction regime and producing GeV-scale electrons and photons exceeding $500\,\mathrm{MeV}$. The resulting $\gamma$-ray emission is directional, peaking around $\pm 20^\circ$, with laser-to-photon energy conversion efficiencies reaching up to $20\%$. These results underscore the importance of self-consistent channel dynamics and instabilities in shaping radiation output in realistic laser-plasma configurations. Unlike approaches relying on preformed channels, the self-organized nature of the cavity enables robust field amplification and high photon yield even under moderate focusing conditions. The interplay between relativistic transparency, radiation reaction, and dynamic cavity formation provides a scalable and experimentally accessible pathway to ultra-bright, high-efficiency $\gamma$-ray sources in the multi-petawatt regime.

This work is supported by the project “Advanced research using high intensity laser produced photons and particles” (ADONIS) \allowbreak{(CZ.02.1.01/\allowbreak0.0/0.0/16 019/0000789)} from the European Regional Development Fund. C.P.R. acknowledges support from EPSRC grant EP/V049461/1. The EPOCH code is in part funded by the UK EPSRC grants EP/G054950/1, EP/G056803/1, EP/G055165/1 and EP/M022463/1.




\bibliographystyle{apsrev4-2}
\bibliography{biblio_2025PRL}

\end{document}